\documentclass[12pt]{article}
\usepackage{epsfig}

\newcommand{\lessim}{\raisebox{-0.8mm}%
{\hspace{1mm}$\stackrel{<}{\sim}$\hspace{1mm}}}
\newcommand{\alphas}{\alpha_{\mathrm{s}}}

\newcommand{\pT}{p_{\perp}}
\newcommand{\pTo}{p_{\perp 0}}
\newcommand{\pTmin}{p_{\perp\mathrm{min}}}
 
\renewcommand{\d}{{\mathrm d}}
\newcommand{\g}{{\mathrm g}}
\newcommand{\p}{{\mathrm p}}
\newcommand{\q}{{\mathrm q}}
\newcommand{\pbar}{\overline{\mathrm p}}
\newcommand{\qbar}{\overline{\mathrm q}}

{\end{list}}
\newcounter{enumct}
\newenvironment{Enumerate}{\begin{list}{\arabic{enumct}.}%
{\usecounter{enumct}\setlength{\topsep}{0.2mm}%
\setlength{\partopsep}{0.2mm}\setlength{\itemsep}{0.2mm}%
\setlength{\parsep}{0.2mm}}}{\end{list}}
 
\newlength{\abstwidth}
\begin{document}
 
\sloppy

\pagestyle{empty}
 
\begin{flushright}
LU TP 00-47\\
hep-ph/0011282\\
November 2000
\end{flushright}
 
\vspace{\fill}

\begin{center}
{\LARGE\bf A Toy Model of Colour Screening}\\[3mm]
{\LARGE\bf in the Proton}\\[10mm]
{\large Johann Dischler and Torbj\"orn Sj\"ostrand}\\ [2mm]
{\it Department of Theoretical Physics,}\\[1mm]
{\it Lund University, Lund, Sweden}
\end{center}

\vspace{\fill}
 
\begin{center}
{\bf Abstract}\\[2ex]
\begin{minipage}{\abstwidth}
In hadronic collisions, the mini-jet cross section is formally
divergent in the limit $\pT \to 0$. We argue that this divergence
is tamed by some effective colour correlation length scale of the 
hadron. A toy model of the hadronic structure is introduced, that 
allows an estimate of the screening effects, and especially their
energy dependence.
\end{minipage}
\end{center}
 
\vspace{\fill}
 
\clearpage
\pagestyle{plain}
\setcounter{page}{1}

The perturbative parton--parton interaction cross section in a 
hadronic collision is divergent roughly like 
\begin{equation}
\frac{\d\hat{\sigma}}{\d\pT^2} \propto \frac{\alphas^2(\pT^2)}{\pT^4} 
\end{equation}
for $\pT \to 0$. This is to be convoluted with parton distributions
$f_i(x,Q^2 \approx \pT^2)$ to give a hadronic interaction cross section. 
The dominant contributions come from scattering by $t$-channel
gluon exchange. With an artificial lower cut-off scale $\pTmin$,  
\begin{equation}
\sigma_{\mathrm{int}}(\pTmin) = \sum_{i,j,k,l} \int \d x_1 \, f_i(x_1, Q^2)
\int \d x_2 \, f_j(x_2, Q^2) \int_{\pTmin^2} \d\pT^2 \,
\frac{\d\hat{\sigma}_{ij \to kl}(\hat{s} = x_1 x_2 s)}{\d\pT^2} ~.
\end{equation}
The jet/mini-jet cross section is twice this, since each scattering gives 
two jets in the lowest-order approximation considered here. Studying e.g. 
Tevatron collider energies, $\sqrt{s} \approx 2$~TeV, 
$\sigma_{\mathrm{int}}(\pTmin)$ exceeds the total $\p\pbar$ 
cross section $\sigma_{\mathrm{tot}}$ for $\pTmin \lessim 3$~GeV. 
Since 3~GeV still is well above the $\Lambda_{\mathrm{QCD}}$ scale 
of $\sim 0.2$~GeV, there is no obvious reason why perturbation theory 
should have broken down, and one would seem to be in trouble. The 
resolution of this paradox probably comes in several steps, and at 
least in two.
 
First of all, the mini-jet cross section above is inclusive.
Thus, if an event contains two parton--parton interactions 
\cite{twoint}, it counts twice in $\sigma_{\mathrm{int}}(\pTmin)$ 
but only once in $\sigma_{\mathrm{tot}}$. Thereby 
$\sigma_{\mathrm{int}}(\pTmin) > \sigma_{\mathrm{tot}}$
becomes allowed. Multiple 
parton--parton interactions is the concept that, based on
the composite nature of hadrons, indeed several parton pairs may scatter
in a typical hadron--hadron collision \cite{maria}. Over the years, 
evidence for this mechanism has accumulated \cite{oldmultint}, such 
as the direct observation by CDF \cite{cdfmultint}. The 
events studied experimentally, with two parton pairs at reasonably 
large $\pT$, only form the tip of the iceberg, however. One may expect 
that most interactions are at lower $\pT$, where they do not produce
visible jets, but only contribute to the underlying event structure. 
As such, they are then believed to be at the origin of a number of key 
features, like the broad multiplicity distributions, the significant 
forward--backward multiplicity correlations, and the pedestal effect 
under jets \cite{maria}.

While the effects of multiple interactions on event properties are 
smaller for lower-$\pT$ scatterings, in the models we studied so far 
the interaction rate increases faster with decreasing 
$\pT$ than the effect per scattering, so there is no stability in the 
limit $\pTmin \to 0$. 

The second necessary aspect is likely that a regularization of the jet 
cross section should occur at small $\pT$ from the fact that the incoming
hadrons are colour singlets --- unlike the coloured partons assumed in 
the divergent perturbative calculations --- and that therefore the 
colour charges should screen each other in the $\pT \to 0$ limit.
Thus $\pTmin$ could take on a physics meaning, roughly, as the 
inverse of some colour screening length in the hadron. Of course, 
one would not expect a sharp cut-off of the mini-jet cross section at 
$\pTmin$, but rather a smooth dampening. Nevertheless the $\pTmin$
parameter provides a useful first approximation. Fits to data typically 
give $\pTmin \approx 2$ GeV. 

One key issue is the energy dependence of $\pTmin$; this may be 
relevant e.g. for comparisons of jet rates at different Tevatron 
energies, and even more for any extrapolation to LHC energies. 
The question actually is more pressing now than at the time of the 
study in \cite{maria}, since nowadays parton distributions are 
known to be rising more steeply at small $x$ than the flat $xf(x)$ 
behaviour normally assumed for small $Q^2$ before HERA. This 
translates into a more dramatic energy dependence of the 
multiple-interactions rate for a fixed $\pTmin$, and thence 
to a charged multiplicity rising faster 
than data \cite{UA5}. Based on such considerations, the 
$\pTmin$ in the {\sc Pythia} program \cite{pythia} was made
explicitly energy-dependent some time ago:
\begin{equation}
\pTmin = (1.9~{\mathrm{GeV}}) \left(
\frac{s}{1~\mathrm{TeV}^2} \right)^{\epsilon} ~,
\label{pomeronpT}
\end{equation}
with $\epsilon = 0.08$. This value is picked to agree with
the $s^{\epsilon}$  behaviour assumed in the parameterization
of the total cross section in hadron--hadron collisions \cite{DL} 
that, via reggeon phenomenology, should relate to the behaviour of 
parton distributions at small $x$ and $Q^2$. A study of the
energy dependence of $\d N_{\mathrm{charged}} / \d\eta |_{\eta=0}$
also confirms that an $\epsilon$ in this range gives sensible
agreement with data \cite{LHCb}.

In the following, we will develop an extremely simple model for the 
proton structure, where the amount of screening may be studied 
\cite{johann}. Our basic approach is to assign an exclusive parton 
configuration to each proton. This configuration is --- for a given 
evolution scale $Q$ --- defined in terms of the longitudinal momentum 
fraction $x$ values, the transverse coordinate positions and colour 
charges of `all' the partons present in a proton. The traditional 
inclusive parton-distribution picture is obtained as the average over 
many such proton `snapshots'. A simple plane wave represents the 
exchange of a gluon between two colliding hadrons. Longitudinal 
coordinates of the partons are not specified, so the exchanged gluon
is only characterized by its transverse momentum. The exclusive proton 
picture then allows a comparison of a coherent versus an incoherent sum 
of colour charges coupling to the plane wave. It is this ratio, coherent 
over incoherent, that will provide our measure of screening between 
partons in the proton. In particular, the long wavelength limit
then explicitly corresponds to a vanishing of interactions between
two colourless hadrons. As the collision energy is increased, more
partons at smaller $x$ values become accessible and the screening 
effects increase in importance.
 
We certainly know this ansatz to be wrong in many of its 
details, but still hope that it will catch enough of the spirit to 
provide some insight. It is clear that a full-fledged description 
--- e.g. with lattice QCD calculations --- is way beyond the 
current capability, so toy models is all that could be offered today.
The approach we present here should be viewed as only one 
possible aspect or formulation of the required dampening 
effect. The vanishing net colour charge of hadrons
has been used in models since long \cite{Low}, as has the spatial size 
of the colour singlet wave function \cite{GS}. Our model also has some 
similarities with the dense-packing view of the proton \cite{GLR}, 
although without the possibility of parton recombinations.
Among alternatives, one could mention the possibility of a nontrivial 
structure of the QCD vacuum \cite{LN}, the solution of a Dyson-Schwinger
equation for the gluon propagator \cite{CR}, the introduction of an
effective gluon mass from lattice QCD results \cite{MO}, and the 
possibility of several hard scatterings within a single parton chain 
\cite{JL}, e.g. formulated in terms of non-integrated structure functions
and the Linked Dipole Model description of non-$\pT$-ordered chains
\cite{GM}. 
 
\begin{figure}[t]
\begin{center}
\rotatebox{270}{\mbox{\epsfig{file=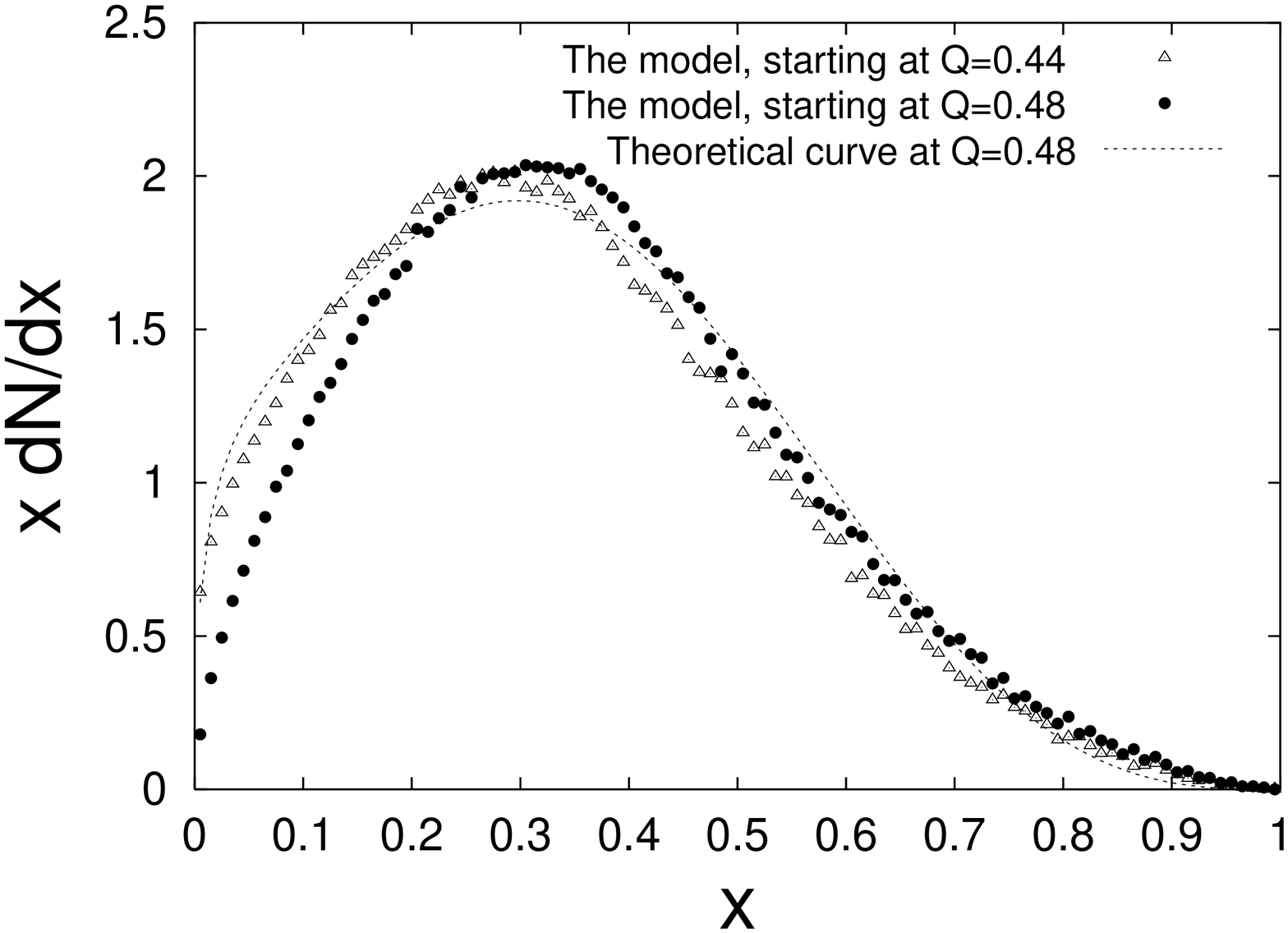, height=8cm}}}%
\rotatebox{270}{\mbox{\epsfig{file=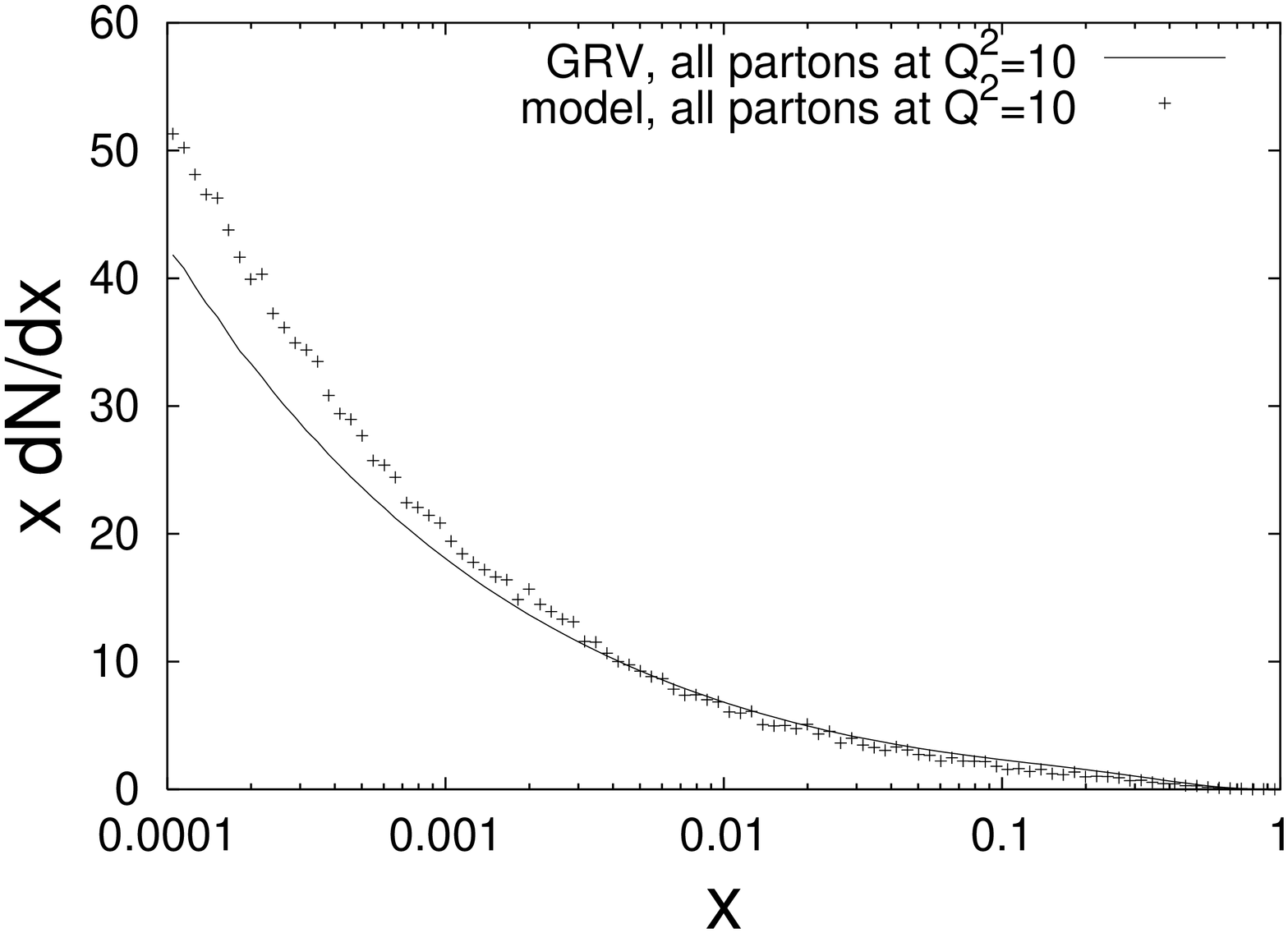, height=8cm}}}
\end{center}
\caption{Parton momentum distribution, summed over quarks and gluons.
Our simulation is compared with the GRV parameterization \cite{GRV}.
(a)  At the small scale $Q_0 = 0.48$~GeV. 
(b) Evolved to $Q^2=10$~GeV$^2$, with a cut so that branchings do 
not produce partons below $x_{min}=10^{-4}$.
\label{fig:xevolve}}
\end{figure}

In our model, the $Q^2$ evolution of the parton distributions is given 
by the conventional DGLAP equations \cite{DGLAP}. As usual, the 
starting configuration requires nonperturbative input. A valence-like 
ansatz at a small $Q_0$ scale \cite{constit} limits the number of
required free parameters. We have therefore used the GRV distributions
as a reference \cite{GRV}, where the three valence quarks together with
two `valence gluons' almost completely define the proton at the small 
scale $Q_0 = 0.48$~GeV. In our simulation, these five partons are first
selected according to a single-particle distribution 
\begin{equation}
f(x) \propto x^{\alpha} (1-x)^{\beta} ~,
\end{equation}
with tuned parameters $\alpha = -0.4$ and $\beta = 1.2$. Thereafter 
all five $x_i$ are normalized so that their sum corresponds to unity:
\begin{equation}
(x_{i})_{\mathrm{norm}}=\frac{x_i}{\sum_{j=1}^5 x_j} ~.
\end{equation}
This means that the original distribution is pushed towards the middle,
and it is especially difficult to tune the low-$x$ tail, at least with 
the simple $f(x)$ ansatz above. A reduction of the $Q_0$ scale from 
0.48 to 0.44 GeV helps improve the small-$x$ behaviour,
Fig.~\ref{fig:xevolve}a. 

From there on, the evolution follows the standard leading-order 
DGLAP formalism, expressed in terms of Monte Carlo-generated 
branchings of partons, $\q \to \q\g$,  $\g \to \g\g$ and
$\g \to \q\qbar$. Unfettered, such an approach gives an infinity of
gluons in the $x \to 0$ limit. For the exclusive approach we have in 
mind, this is not so physical. In the study of scatterings above
some $\pTmin$ scale, only partons above some related $x_{\mathrm{min}}$ 
scale can at all interact. Partons below 
$x_{\mathrm{min}}$ then have a dubious existence. To simplify the 
picture, we assume they are not resolved, i.e. that only branchings 
producing both daughters above $x_{\mathrm{min}}$ are allowed.
This reduces the amount of evolution, but not as dramatically as
might be guessed at first glance. In Fig.~\ref{fig:xevolve}b we see that,
in a typical case, the parton density with an $x_{\mathrm{min}}$ 
constraint is at most 20\% above the no-constraints normal evolution 
in the region $x > x_{\mathrm{min}}$, while the former obviously 
vanishes for $x < x_{\mathrm{min}}$. Disagreements are more visible 
--- but also less interesting for us --- in individual parton species, 
especially for sea quarks, which are not part of our simple ansatz at 
$Q_0$ but are in the GRV one. 

The kinematics of a hard scattering requires that
$x_1 x_2 = \hat{s}/s \geq 4\pT^2/s$. For collisions at central 
rapidities this suggests $x_{\mathrm{min}} \simeq 2\pT/E_{\mathrm{CM}}$.
Away from the center, $x_{\mathrm{min}}$ would be smaller on one side,
down to $4\pT^2/E_{\mathrm{CM}}^2$, and correspondingly larger on the 
other, which to some extent should cancel out. Therefore we will have 
as standard scenario that $x_{\mathrm{min}}$ should increase 
proportionately to the $\pT$ considered, but will include as alternative 
a fixed $x_{\mathrm{min}}$. Furthermore, we will make the association
that $E_{\mathrm{CM}} \propto 1/x_{\mathrm{min}}$, for $x_{\mathrm{min}}$
evaluated at some fixed reference $\pT$ like 1~GeV, but allow as 
alternative that the relation could be more like 
$E_{\mathrm{CM}} \propto 1/\sqrt{x_{\mathrm{min}}}$.

Next we consider the assignment of transverse position coordinates.
The original five partons are here selected according to a Gaussian
shape, with a radius of 0.7~fm in each of the two transverse
dimensions. This gives a minimal correlation between the position
coordinates of these five partons, and thus probably errs on the side 
of simplicity. Attempts with somewhat more correlated forms gave 
fairly similar results, however, so the Gaussian ansatz is not so
critical.

The spatial extent of a fluctuation like $\q \to \q\g \to \q$,  
$\g \to \g\g \to \g$ and $\g \to \q\qbar \to \g$ is of order $1/Q$
according to the uncertainty relation. Therefore, when a DGLAP
branching occurs at a scale $Q$, the two daughters are assumed
to have time to fluctuate a distance of this order away from their
production coordinates before they branch in their turn, or are probed
by our plane gluon wave. The daughter partons are assumed to move out 
in opposite directions from the production vertex, with a uniform 
azimuthal distribution. The two distances are picked 
independently of each other, uniformly between 0 and $1/Q$, 
alternatively between 0 and $2/Q$. We will see that this choice is
relevant, whereas results are essentially equivalent 
e.g. for exponentially dampened distributions with the same mean.
For a parton which takes the momentum fraction $z$ in a branching, 
an additional factor $\sqrt{1-z}$ is included, giving an approximate 
$\pT/Q$ dampening factor for the separation in transverse rather than 
longitudinal direction. Here the relation $\pT = \sqrt{1-z} Q$ follows
in light-cone kinematics assuming the mother and recoiling partons to be 
massless. It is of some relevance that a very soft gluon emission does 
not affect the position of the hard parton unduly, since else multiple
soft gluons could lead to a too rapid increase of the proton radius with
$Q$. For the final result here, however, the $\sqrt{1-z}$ factor turns 
out not to be crucial.

Finally we come to the colour space picture. Here we have picked a 
simple planar representation, with the three primary colours placed
in a triangle around the origin.  Thus the full phase information is lost; 
i.e. it is possible to ensure that the proton state is colour neutral but 
not that it is in a singlet. This certainly is a major simplification of 
the real world, and one which it is difficult to estimate the impact of. 
A gluon emission corresponds to a quark changing colour, e.g.
$\q(r) \to \q(b) + \g(r\overline{b})$. The emitted gluon thus carries one 
colour and a different anticolour, i.e. the two colour diagonal gluon 
states are not populated; another simplification. Further, the ratio of 
gluon to quark charge is $\sqrt{3}$ versus $\sqrt{N_C/C_F} = 3/2$ in QCD.  
Also in a gluon branching to two, a new colour is picked, e.g.
$\g(r\overline{b}) \to \g(r\overline{g}) + \g(g\overline{b})$, in order
to avoid diagonal gluons. The gluon branching to quarks is unambiguous,
$\g(r\overline{b}) \to \q(r) + \qbar(\overline{b})$. 
For colour assignments in the starting configuration, the three valence 
quarks are first picked to be $r + g + b$, and then the two gluons are 
handled as if emitted from two of the quarks, picked at random.

An effort is thus made to retain overall vanishing colour for all partons 
in a proton that are seen by an exchanged gluon. In case of multiple 
interactions, it could be argued that only the `first' exchange would be
between singlet hadrons, while `subsequent' ones would be between already
coloured hadrons. Clearly, the imposition of such a time ordering can be
questioned but, even if taken at face value, colour screening could still
provide a significant dampening of the naive perturbative answer.
   
To summarize, an explicit parton configuration of the proton is constructed
for any value of the evolution parameter $Q$ above some low $Q_0$ starting
scale. Each parton $k$ is characterized by its momentum fraction $x_k$, its
transverse position coordinate $\mathbf{r}_k$ and its two-dimensional
colour charge $q_k$. If such a proton is probed by a gluon plane wave function
of transverse momentum $\mathbf{p}_{\perp}$, we may define a ratio
\begin{equation}
A = \frac{|\sum_k q_k e^{i \mathbf{r}_k \mathbf{p}_{\perp}}|^2}{\sum_k |q_k|^2} 
\label{Adef}
\end{equation}
between a coherent and an incoherent sum of the colour charges. Note that
we assume the exchanged gluon to be a superposition of all possible
colour-anticolour pairs, so that we do not need to single out a special 
direction in colour space.  

\begin{figure}[t]
\begin{center}
\rotatebox{270}{\mbox{\epsfig{file=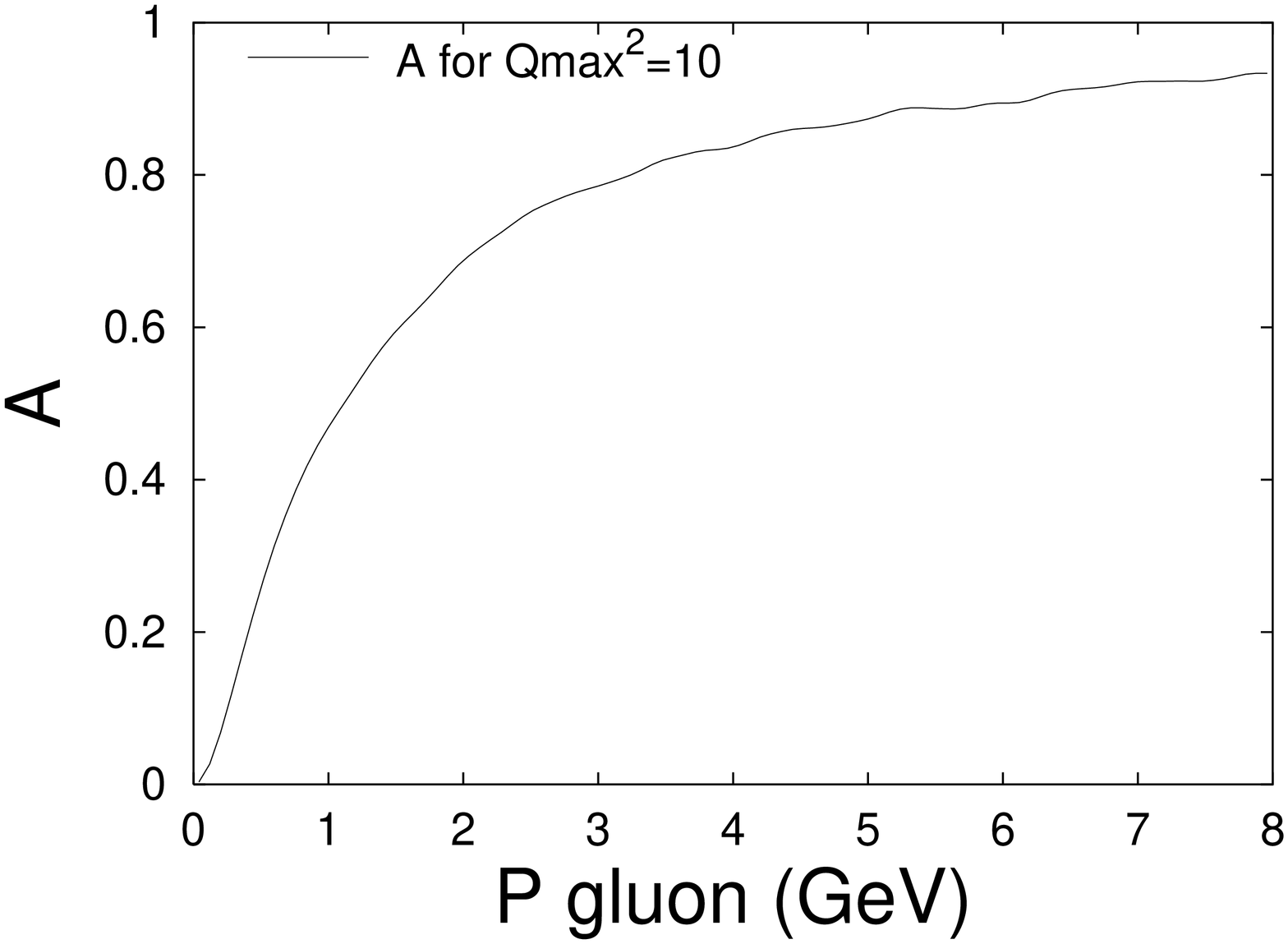, height=8cm}}}%
\rotatebox{270}{\mbox{\epsfig{file=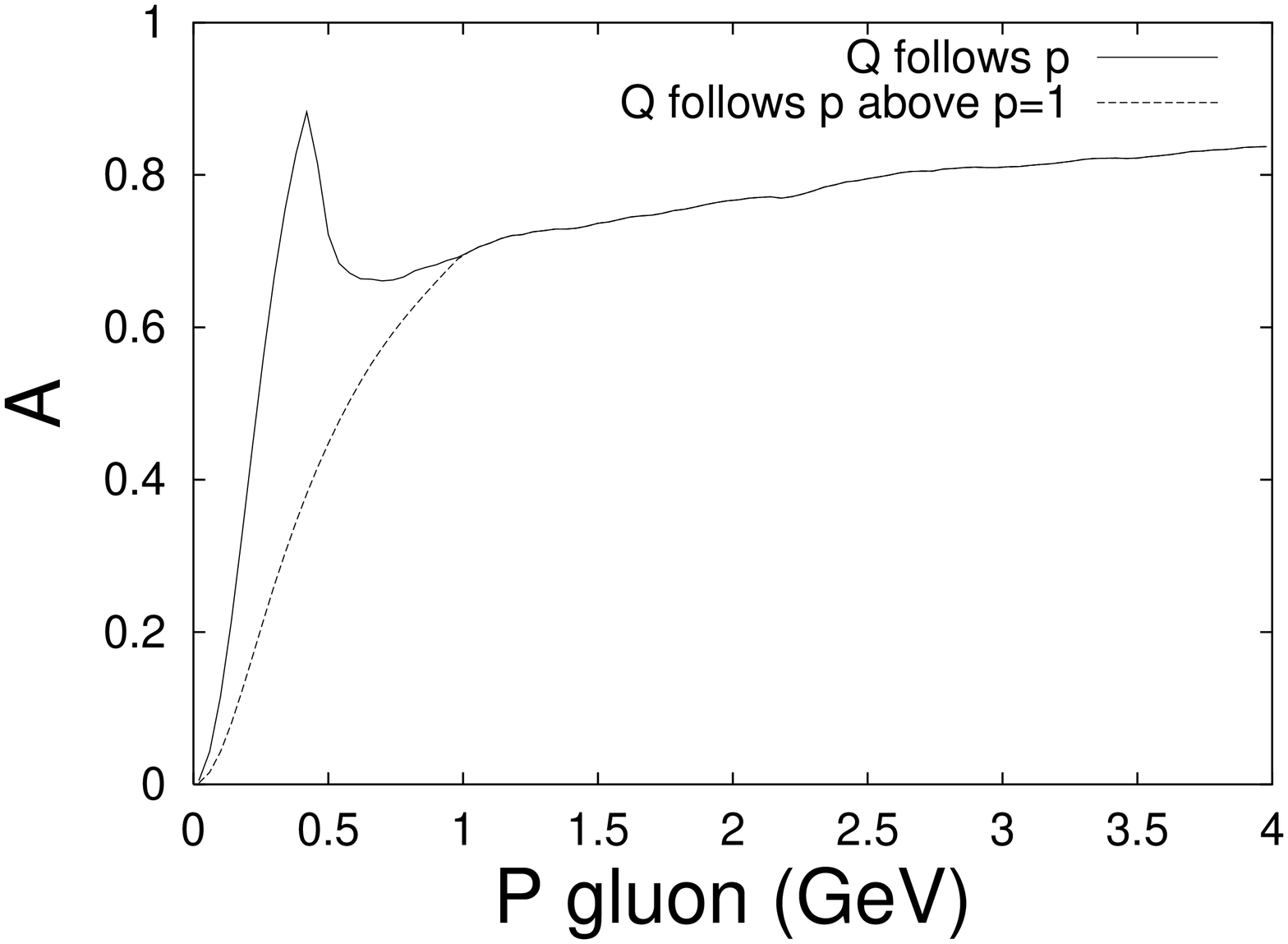, height=8cm}}}\\[3mm]
\hspace*{\fill} (a)\hspace*{\fill} \hspace*{\fill} (b) \hspace*{\fill} \\
\rotatebox{270}{\mbox{\epsfig{file=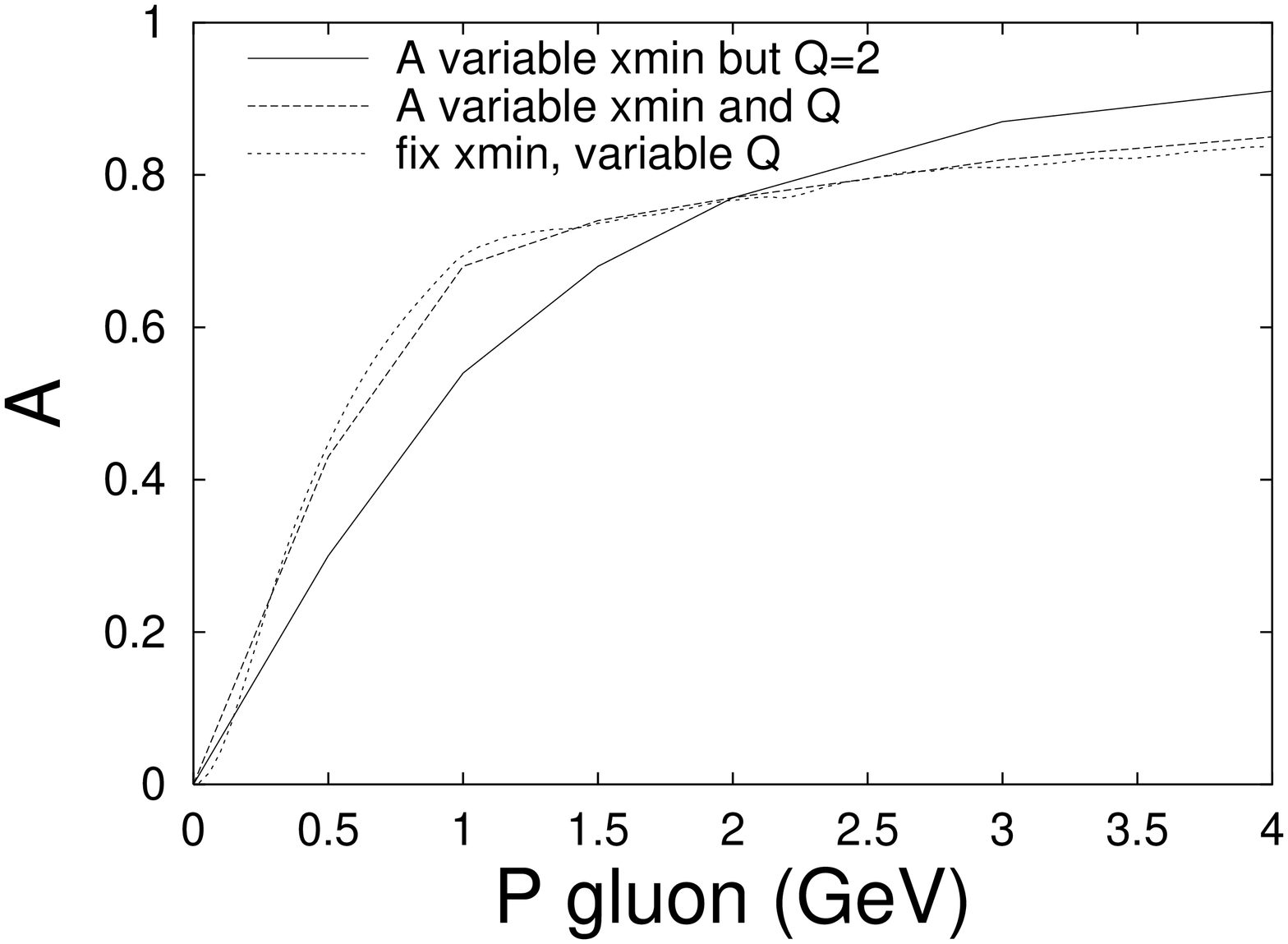, height=8cm}}}%
\rotatebox{270}{\mbox{\epsfig{file=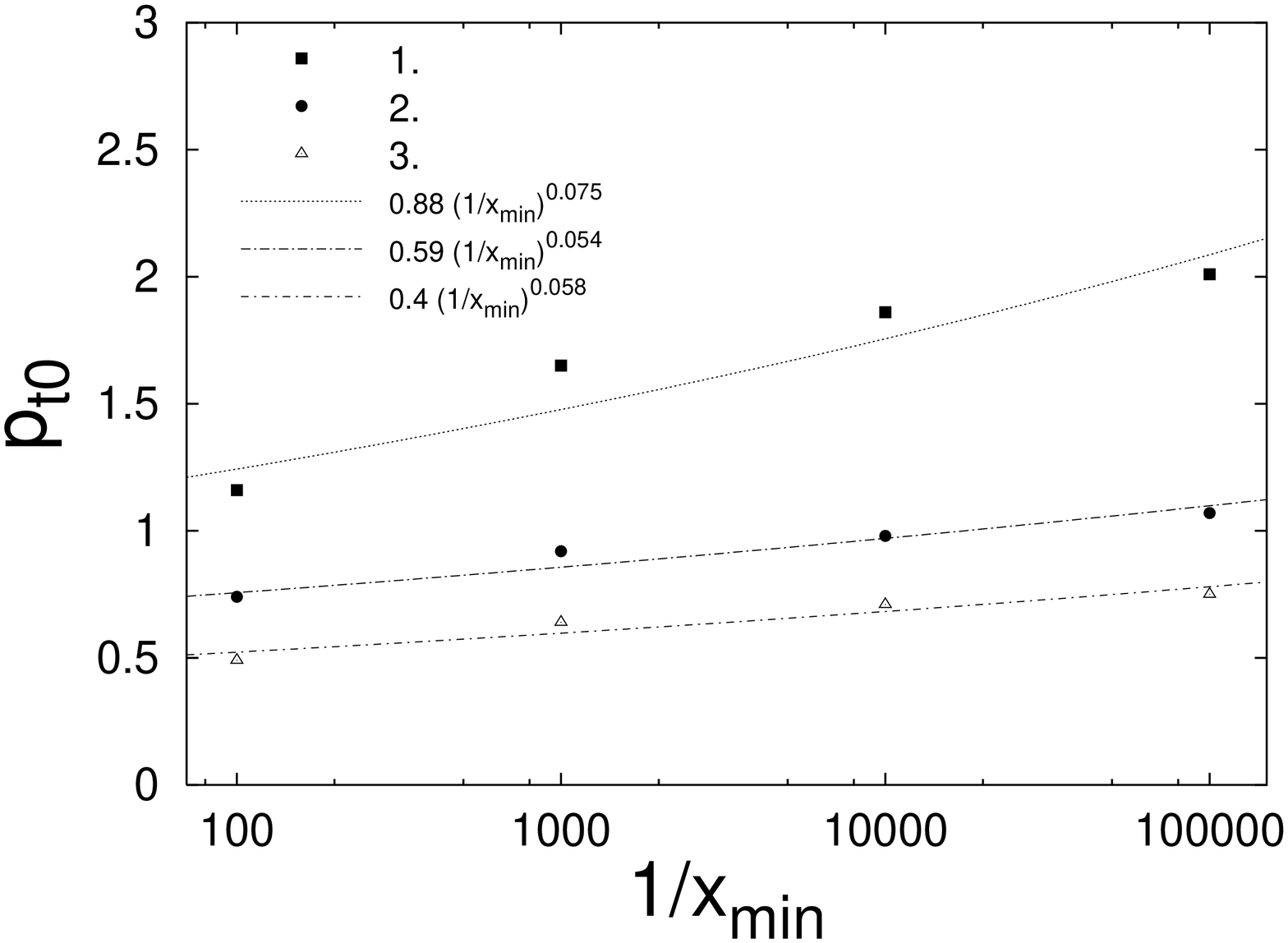, height=8cm}}}\\[3mm]
\hspace*{\fill} (c) \hspace*{\fill} \hspace*{\fill} (d) \hspace*{\fill}
\end{center}
\caption{(a-c) The amount of screening, $A$ in eq.~\protect\ref{Adef}, 
as a function of the probing gluon (transverse) momentum $P = \pT$.
(a) For a proton at fixed evolution scale $Q^2 = 10$~GeV$^2$, and 
fixed $x_{\mathrm{min}} = 10^{-3}$.
(b) For a proton where the proton evolution scale $Q = \pT$ for
$\pT \geq 0.44$~GeV, alternatively 1 GeV, again with 
$x_{\mathrm{min}} = 10^{-3}$.
(c) Full: variable $x_{\mathrm{min}} = 10^{-3}\pT$ and fixed $Q=2$~GeV.
Dashed: variable $x_{\mathrm{min}} = 10^{-3}\pT$ and 
$Q = \max(\pT, 1~\mathrm{GeV})$.
Dotted: fix $x_{\mathrm{min}} = 10^{-3}$ and variable
$Q = \max(\pT, 1~\mathrm{GeV})$.
(d) The $x_{\mathrm{min}}$ dependence of the $\pTo$ effective cut-off.
See text for a explanation of the three scenarios.
\label{fig:screen}}
\end{figure}

The screening effect is most easily observed if one fixes the $Q$ of the
probed proton and then varies the $\pT$ of the probing gluon, 
Fig.~\ref{fig:screen}a. The suppression factor $A$ then has to vanish 
quadratically as $\pT \to 0$, but in the interesting $\pT$ range
the dominant aspect is instead the slower rise to the asymptotic
limit $A = 1$. The choice of a fixed $Q$ scale is representative
for the way parton distributions de facto are used with $Q_0$ as scale 
for $Q < Q_0$. For popular distributions like the current CTEQ and 
MRST ones \cite{currpdf}, where $Q_0^2 \geq 1$~GeV$^2$, 
the inclusion of a suppression factor then significantly modifies the
picture for $Q < Q_0$. Actually $Q_0$ has tended to come down with time;
some years ago $Q_0^2 \geq 4$~GeV$^2$ was the norm. Also remember that 
the cross section suppression is $A^2$, i.e. with one factor of $A$ for 
each hadron beam.

A conventional choice, but not a unique one, is to pick the $Q$ scale of 
parton distributions to agree with the $\pT$ scale of the hard interaction. 
Then the approach $A \to 1$ is much slower at large $\pT$, since the
branchings that occurred at scales $Q$ not so much smaller than $\pT$
are not fully resolved, Fig.~\ref{fig:screen}b. We remind that the proton 
seen by a gluon probe is a much more `busy' place than the one seen by 
a photon in Deeply Inelastic Scattering processes, where the gluon content 
is not probed directly. Thus the destructive interference we introduce 
here does not invalidate the conventional structure function measurements 
and partonic interpretations. In hadronic physics, higher order 
perturbative corrections could partly contain, and partly mask, the
physics of an $A$ somewhat below unity.

More interesting is the behaviour at small $\pT$. We here note a spike 
in $A$ at the scale $\pT = Q_0 = 0.44$~GeV, Fig.~\ref{fig:screen}b.
At this scale, the proton consists of the original five partons,
and these are normally well separated in space. Above it, the 
partons emitted in the evolution process tend to have a more clustered 
spatial distribution and therefore screen more. Below it, there is no  
further evolution of the proton but only a straightforward destructive
interference. Thus the spike is an artifact of how we discontinuously 
in $Q$ go from a fixed proton picture to a very rapid evolution; remember 
that the evolution rate behaves like $\alphas(Q^2) \, \d Q^2/Q^2$ and thus
is as largest just at the lower cut-off. In order to provide a smoother
and physically maybe more sensible picture, we therefore do not probe the
proton at scales below 1~GeV, i.e. put $Q = \max(\pT, 1~\mathrm{GeV})$.
The original $Q_0$ scale and evolution below 1~GeV still lives on in
terms of a sensible $x$/$\mathbf{r}$/colour configuration at 1~GeV.

Fig.~\ref{fig:screen}b is based on a fixed $x_{\mathrm{min}}$ scale. 
However, an $x_{\mathrm{min}}$ scaling $\propto \pT$ behaves very similarly,
Fig.~\ref{fig:screen}c, if the two are matched for $\pT=1$ GeV.
If instead $Q$ is fixed, the shape is more changed.

We now turn to the prime objective of this paper, namely to study 
the energy dependence of $\pTmin$. Actually, the sharp 
cut-off $\theta(\pT - \pTmin)$ is not very physical. A more likely 
behaviour is a dampening factor something like $\pT^2/(\pT^2 + \pTo^2)$,
where $\pTo$ takes over the role of free parameter. With one such
factor per incoming proton, the $1/\pT^4$ singularity of the perturbative
parton-parton cross section is regularized. 

Whereas the proposed dampening form does not completely reproduce the 
curves shown above, qualitatively there is agreement. Instead of doing  
a fit to the curve shape, $\pTo$ is extracted as the $\pT$ value for 
which $A = 1/2$. This would also have been a sensible definition of a 
$\pTmin$, backed up by experience with the impact of the two regularization
procedures on fits to data, where $\pTo \approx \pTmin$ is obtained 
\cite{maria}.

The resulting $\pTo$ values are plotted as a function of 
$1/x_{\mathrm{min}}$ in Fig.~\ref{fig:screen}d. Three scenarios are compared:
\begin{Enumerate}
\item The proton probed at a fixed $Q=2$~GeV scale, travel distance of a 
parton is distributed between 0 and $1/Q$ of the branching where it is 
produced. 
\item Also proton probed at a fixed $Q=2$~GeV scale, but travel distance 
twice as large.
\item The proton is probed at a running $Q = \pT$, travel distance
between 0 and $2/Q$ as in 2. 
\end{Enumerate}
In all these three cases, representative for several more studied by us,
a clear energy-dependence of $\pTo$ is observed.
If one tries to make a fit to an exponential form 
$\pTo \propto 1/x_{\mathrm{min}}^{\delta}$, typically one obtains
a $\delta$ in the range 0.05 to 0.08. However, two comments are in order.
One is that the curves tend to rise less steeply at higher energies than
implied by the fit form. The other is that, if one attempts the association  
$1/x_{\mathrm{min}} \propto E_{\mathrm{CM}}$, then eq.~(\ref{pomeronpT})
would have predicted $\delta = 2\epsilon = 0.16$, i.e. a much steeper slope.
The alternative association $1/x_{\mathrm{min}} \propto E_{\mathrm{CM}}^2$ 
would give nicer agreement, but we iterate that it may be the less plausible
ansatz.

In summary, we have presented a very simpleminded model of the longitudinal 
momentum, transverse coordinates and colour structure of the proton, in order 
to be able to estimate colour screening effects at small $\pT$. Many of the
details turn out to be less critical, but there are two key assumptions.
The first is that the proton, by virtue of being a colour singlet, should couple
to a soft gluon with reduced strength. The second is that, at higher energies, 
the proton can effectively be resolved into more partons that, in principle, 
can interact to give scatterings at a given $\pT$ scale. We show that these two 
assumptions lead to the conclusion that the perturbative ansatz for the cross
section is only valid above some $\pTmin$ scale that increases with energy.
If this increase is fitted to a form 
$\pTmin \propto E_{\mathrm{CM}}^{\kappa}$, any value $\kappa$ in the range
0.05 to 0.16 could be obtained, which certainly overlaps with an experimentally
acceptable behaviour. However, one lesson is that the ansatz in 
eq.~(\ref{pomeronpT}) may tend to overestimate the rate of increase at large 
energies, and thereby e.g. lead to an underestimate of multiplicities at LHC 
energies. Conversely, the study of minimum bias events at LHC may help improve 
the understanding of very low-$\pT$ physics processes.

\end{document}